\documentclass{article}

\usepackage{PRIMEarxiv}

\usepackage[utf8]{inputenc} 
\usepackage[T1]{fontenc}    
\usepackage{hyperref}       
\usepackage{url}            
\usepackage{booktabs}       
\usepackage{amsfonts}       
\usepackage{nicefrac}       
\usepackage{microtype}      
\usepackage{lipsum}
\usepackage{fancyhdr}       
\usepackage{graphicx}       
\graphicspath{{media/}}     

\usepackage{xcolor}

\title{Heat Transfer in Composite Materials: Mechanisms and Applications
}

\author{
  Mohammad Alaghemandi, Morgan Alamandi\\
  Department of Chemistry, Boston University, Boston, MA, USA \\
  Department of Computer Science, Metropolitan College, Boston University, Boston, MA, USA \\
  \texttt{alaghemandi@gmail.com} \\
}

\begin{document}
\maketitle

\begin{abstract}
Understanding heat transfer in composite materials is essential for optimizing their performance in critical applications across industries such as aerospace, automotive, renewable energy, and construction. This review offers a comprehensive examination of the various heat transfer mechanisms within composite materials and explores how these processes, spanning different length and time scales, are influenced by the materials' composition and structure. Both traditional and advanced analytical and numerical modeling techniques are explored, emphasizing their importance in predicting and optimizing thermal behavior across these scales. Furthermore, the review evaluates current experimental methods for measuring thermal properties, discussing their limitations and potential areas for enhancement.
Significant attention is devoted to the practical applications of composite materials, from thermal management in electronic devices to heat-resistant components in aerospace engineering. Recent innovations, such as the integration of phase change materials and the development of nano-enhanced composites, are assessed for their potential to transform heat transfer capabilities. Ongoing challenges are addressed, and future research directions are outlined, highlighting the need for advancements in material science and engineering to meet emerging demands. This review aims to bridge the gap between fundamental research and practical applications, providing a comprehensive understanding of heat transfer in composite materials that is both rooted in current science and driven by future possibilities.
\end{abstract}


\section{Introduction}
Composite materials, created by combining two or more constituent materials with distinct properties, play a vital role in enhancing performance across diverse applications~\cite{Hsissou2021,Gay2022}. By leveraging the complementary attributes of their individual components, composites achieve unique property profiles that exceed the capabilities of single materials~\cite{clyne2019introduction}. Among these properties, effective thermal management has become increasingly important as technological advancements push devices and systems toward higher power densities, miniaturization, and more demanding operational conditions. This underscores the need for a comprehensive understanding of heat transfer mechanisms in these complex materials~\cite{li2024overview,qi2022promoting}.

A defining characteristic of composites is their adaptability for specific thermal performance requirements. Typically, a matrix material encapsulates reinforcing fibers or fragments, enhancing mechanical strength while reducing weight, and simultaneously playing a pivotal role in thermal conductivity~\cite{Rajak2019}. Thermally conductive fillers, such as carbon nanotubes or graphene, create efficient heat-conducting networks that optimize the dissipation of thermal energy within the composite~\cite{subramaniyan2024microscale,tan2024thermal}. However, polymer matrices, widely employed in composite systems, generally exhibit relatively low thermal conductivity compared to reinforcements, necessitating innovative designs to achieve optimal thermal performance. Furthermore, fibers often display anisotropic heat transfer properties, conducting heat more effectively along their length than across their diameter~\cite{CHOY1977984}. Understanding these complex interactions in heterogeneous systems, where constituents exhibit varying thermal behaviors, is fundamental to designing advanced composites~\cite{lebeda2024shaping}.
In addition, efficient thermal energy management is vital for maintaining the structural integrity and functionality of composites under extreme conditions. For example, in aerospace applications, composites are exposed to high temperatures and steep thermal gradients during high-speed flight or atmospheric re-entry~\cite{lebeda2024shaping,tan2024thermal}. Similarly, in the automotive sector, components such as brake systems and engine parts rely on materials capable of rapidly dissipating heat to prevent thermal degradation and ensure safety~\cite{qi2022promoting}. In the electronics industry, the miniaturization of devices has intensified the demand for advanced thermal management solutions. Composites are increasingly utilized in heat sinks and thermal interface materials (TIMs) to maintain safe operating temperatures and enhance device reliability~\cite{wei2021thermal,cao2023thermal}. These diverse applications underscore the necessity of tailoring composite materials to meet the thermal control requirements of high-performance systems.

Heat transfer within composites occurs across multiple scales. At the microscale, defined as dimensions less than 10$^{-6}$ meters, thermal properties are governed by the characteristics of the matrix and reinforcements, including thermal conductivity, heat capacity, and resistance. Factors such as the size, shape, and orientation of reinforcements influence thermal pathways, thereby affecting the overall performance of the composite. For instance, high aspect-ratio fillers like carbon fibers or silicon carbide whiskers create continuous conductive networks that promote directional heat flow~\cite{hanus2021thermal,wei2021thermal,cao2023thermal}. Furthermore, a well-bonded interface between the matrix and reinforcement minimizes thermal resistance, while poor adhesion or defects at the interface act as thermal bottlenecks~\cite{tan2024thermal,li2024mastering}. At the macroscale, encompassing dimensions greater than 10$^{-4}$ meters, the composite’s overall architecture plays a pivotal role in thermal performance. Layering or integrating sections with varying thermal properties can be optimized to meet specific application requirements, such as thermal dissipation in electronics or insulation in structural materials~\cite{chernatynskiy2018thermal}.

Despite significant advancements, predicting the thermal behavior of composites remains a challenging task. The intrinsic heterogeneity of these materials often leads to non-uniform heat transfer, which can compromise performance and reliability~\cite{Ma2023}. Conventional homogenization techniques, though widely used, often fail to account for critical factors such as thermal boundary resistance at interfaces, anisotropy due to filler alignment, and structural defects that disrupt heat flow~\cite{subramaniyan2024microscale,li2024overview}. These limitations are particularly pronounced in applications requiring materials to endure extreme temperatures and thermal gradients. Advanced computational approaches that integrate detailed microstructural characterization and multi-scale simulations offer a promising path forward~\cite{li2024mastering,lebeda2024shaping}. These methods, validated against experimental data, are essential for developing materials capable of meeting the evolving demands of high-performance applications~\cite{Dong2022,Zhang2020,Zheng2021,Zha2024}.

As research continues to deepen our understanding of heat transfer in composites, emerging trends emphasize the integration of innovative materials and techniques. For instance, hybrid composites combining organic and inorganic fillers offer opportunities to tailor thermal and mechanical properties for highly specialized applications~\cite{Yue2023ThermoViscoelastic}. Similarly, advancements in additive manufacturing techniques have enabled the creation of composites with precise geometries and controlled filler distribution, further enhancing their thermal performance~\cite{xu2023fiber}. The development of self-healing composites and materials with adaptive thermal conductivity represents another frontier, promising to extend the lifespan and reliability of components under fluctuating thermal loads~\cite{wang2024adaptive}.

This review provides a comprehensive overview of heat transfer mechanisms in composite materials, emphasizing advancements in modeling techniques, experimental methodologies, and practical applications. By synthesizing current knowledge and identifying key research gaps, the review aims to deliver valuable insights for scientists and engineers striving to optimize thermal management in composite materials and bridge the gap between fundamental research and real-world applications.

\section{Fundamentals of Heat Transfer in Composite Materials}

Understanding the various mechanisms through which heat transfer occurs in composite materials is crucial for their design and effective application, especially in industries where thermal management is pivotal~\cite{burger2016review,zhang2020recent,guo2023advances}. These mechanisms include conduction, convection, and radiation, each influenced by factors such as material composition, reinforcement geometry, and environmental conditions~\cite{boyard2016heat}.

\subsection{Conductive Heat Transfer in Composite Materials}
\begin{figure}
    \centering
    \includegraphics[width=0.9\linewidth]{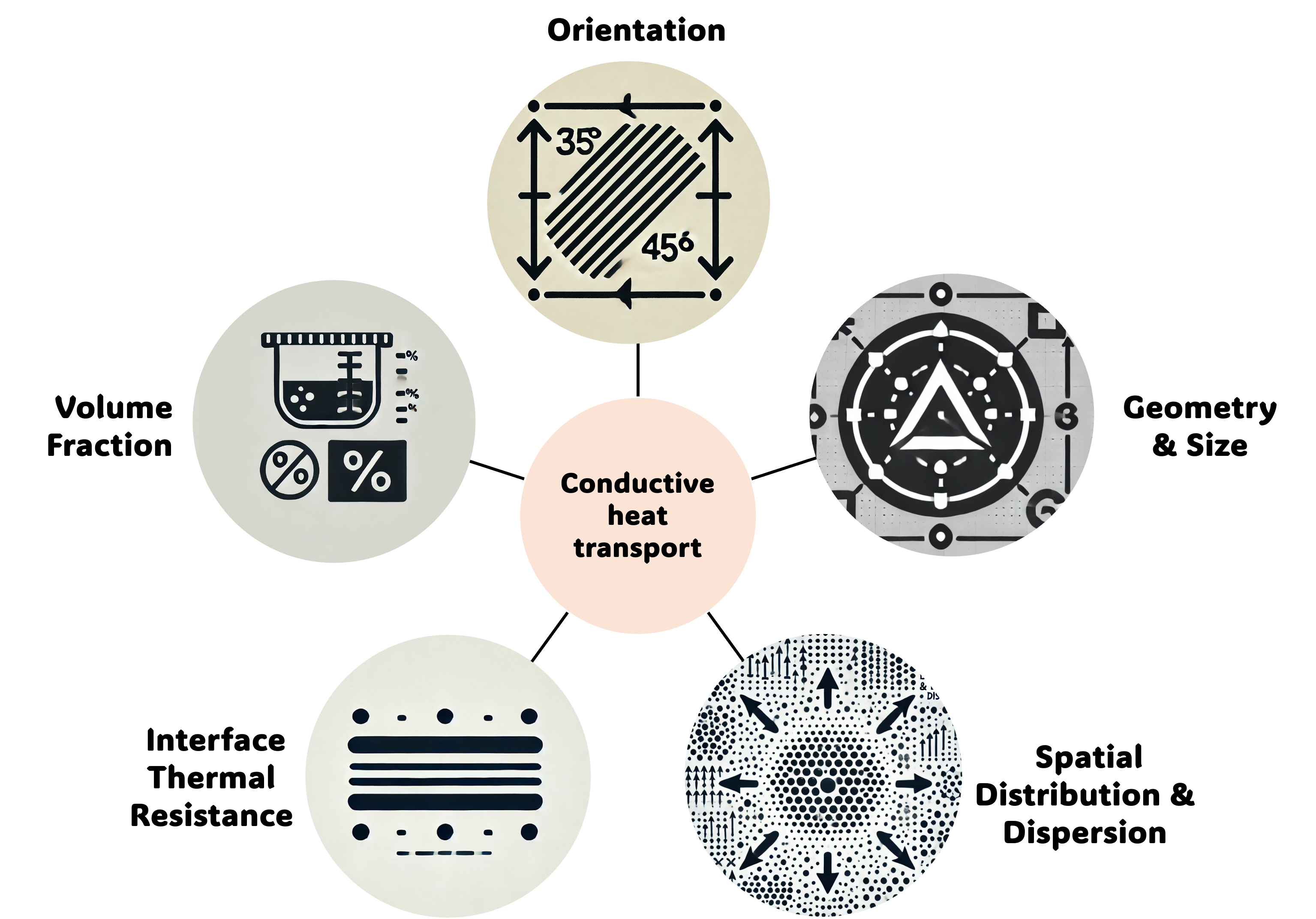}
    \caption{Illustration of factors influencing conductive heat transfer in composite materials, highlighting the effects of reinforcement orientation, filler geometry and size, spatial distribution and dispersion of fillers, interfacial thermal resistance between matrix and reinforcements, and the volume fraction of reinforcement materials.}
    \label{fig:conductive}
\end{figure}

Heat conduction is the primary mechanism of heat transfer in solid composite materials~\cite{kakacc2018heat}. Unlike homogeneous materials, the thermal conductivity of composites is not a straightforward combination of the conductivities of their individual components~\cite{GhaffariMosanenzadeh2022}. Instead, it is significantly influenced by factors such as the properties of the constituents, their volume fractions, geometrical arrangements, and the nature of interfaces between them~\cite{guo2020factors}. Understanding these factors is essential for designing composites with tailored thermal conductivities for specific applications~\cite{tan2024thermal}.
The overall thermal conductivity of a composite material depends on both the intrinsic conductivities of the matrix and reinforcement phases and how these phases are distributed within the composite. Burger et al.~\cite{burger2016review} emphasize that optimizing the orientation, shape, and type of reinforcements within the matrix can significantly enhance the overall thermal conductivity by facilitating more efficient heat flow pathways through the material; see Figure~\ref{fig:conductive}.

\subsubsection{Impact of boundary interface on thermal resistance in composites}
Interfaces between the reinforcement and the matrix in composite materials introduce a significant thermal boundary resistance, also known as Kapitza resistance, which plays a crucial role in heat transfer within these systems~\cite{su2018theory,Starkov2024}. This resistance primarily arises due to the mismatch in the acoustic properties and the strength of bonding between the two phases, which impedes the transfer of phonons across the interface~\cite{gharib2015interfacial,Gao2021}. As phonons are the primary carriers of thermal energy in solids, any disruption in their flow due to interface irregularities can dramatically impact the overall thermal conductivity of the composite~\cite{Ruan2020,Zha2024}.
Minimizing interface thermal resistance is essential for enhancing the thermal performance of composites~\cite{Wei2022}. Improving the interfacial bonding between the matrix and the reinforcement can be achieved through various methods including chemical treatments, the use of coupling agents, or through surface modifications of the fillers~\cite{mohit2018comprehensive,Jasmee2021}. Such enhancements in interfacial bonding facilitate better phonon transport across the interface, thereby reducing thermal resistance and improving the composite's overall thermal conductivity~\cite{gharib2015interfacial,Zhang2024,Wang2024}.

In addition to the mentioned strategies, transcrystallinity at the fiber-matrix interface in thermoplastic composites provides a method for reducing interface thermal resistance~\cite{chen1992effects,quan2005transcrystallinity}. This phenomenon occurs when polymer chains crystallize in an oriented manner around reinforcement fibers, enhancing phononic contact and improving thermal and electrical transport~\cite{farhadpour2024significance}.
Transcrystallinity indicates strong interfacial bonding and serves as a metric to assess fiber-matrix adhesion quality~\cite{wang2020origin,brodowsky2018investigation,raimo2015origin}. Promoting conditions that foster transcrystallinity minimizes interface defects and optimizes thermal conductivity~\cite{looijmans2022synergy}. This is crucial in applications where fiber alignment does not align with heat flow direction, allowing for effective heat removal from the matrix~\cite{qiu2014effect}.

\subsubsection{Influence of reinforcement orientation}
The alignment of fibers or particles within the composite plays a crucial role in determining thermal conductivity~\cite{xu2020thermal,Wang2021}. When fibers are aligned parallel to the direction of heat flow, they provide continuous pathways that facilitate efficient heat conduction~\cite{alaghemandi2011thermal}. This orientation maximizes the contribution of the high-conductivity reinforcement phase to the composite's overall thermal conductivity~\cite{Yi2021}. Conversely, fibers aligned perpendicular to the heat flow can impede heat transfer, as the heat must cross interfaces between the matrix and the fibers more frequently, introducing additional thermal resistance~\cite{reis2011heat,Samal2022}.
In unidirectional fiber-reinforced composites, the thermal conductivity along the fiber direction can be significantly higher than that perpendicular to the fibers, leading to anisotropic thermal properties~\cite{alaghemandi2012thermal}. This anisotropy can be advantageous in applications where directional heat flow is desired. Designing the fiber orientation allows engineers to tailor the thermal conductivity in specific directions to meet the requirements of advanced thermal management systems~\cite{Lebeda2024}.

\subsubsection{Effect of reinforcement geometry and size}
\begin{figure}
    \centering
    \includegraphics[width=0.9\linewidth]{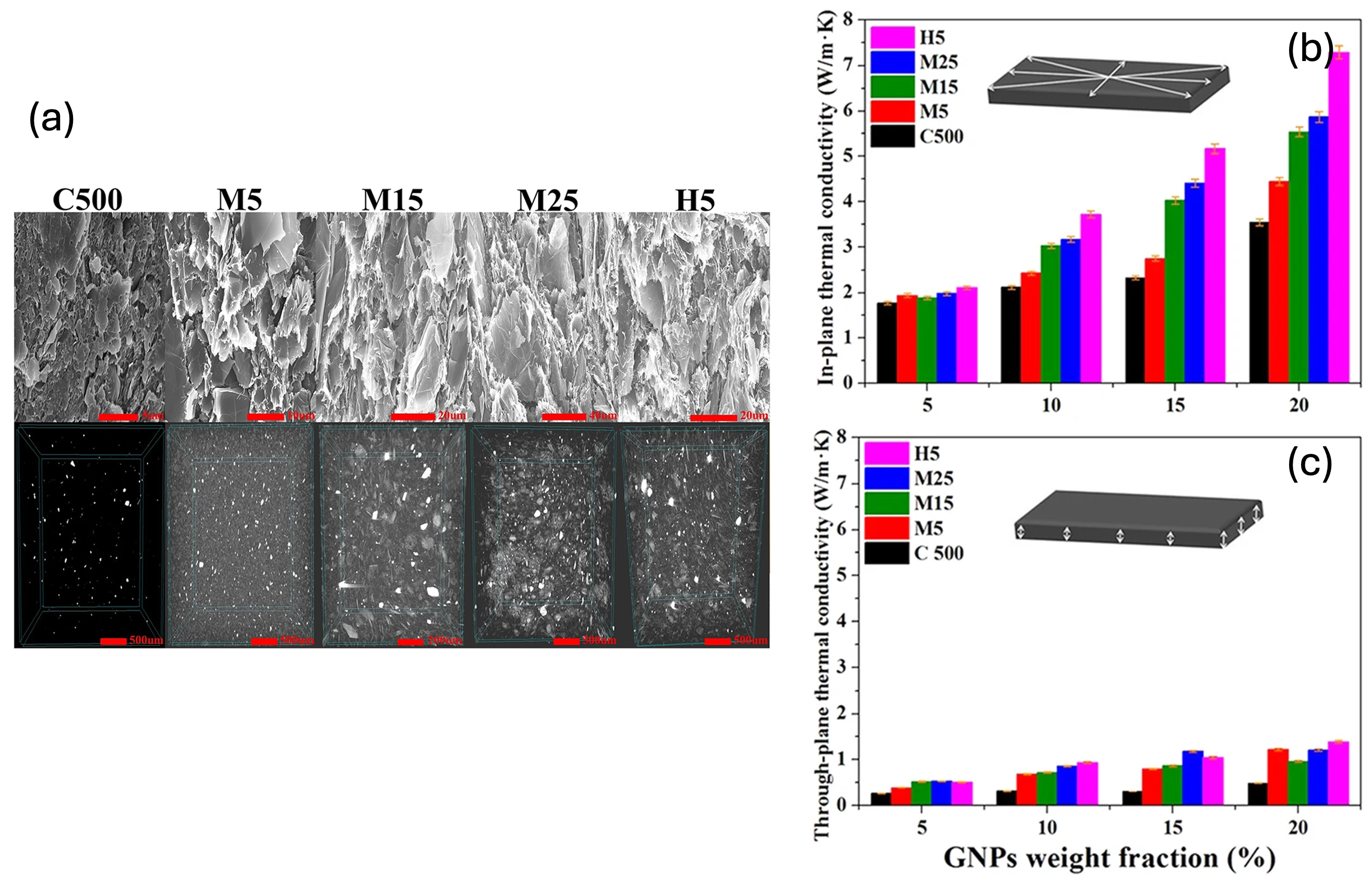}
    \caption{(a) SEM and micro X-ray CT images of polycarbonate (PC) composites filled with 20~wt$\%$ graphene nanoplatelets (GNPs) of varying lateral sizes and thicknesses, illustrating the dispersion and network formation of the fillers. (b) In-plane thermal conductivity and (c) through-plane thermal conductivity of the composites filled with GNPs of different lateral dimensions and thicknesses, demonstrating how the size and thickness of GNPs influence thermal conductivity in both directions~\cite{kim2016thermal}.}
    \label{fig:Kim}
\end{figure}

The geometry and size of reinforcements, such as fibers, particles, or platelets, impact the thermal conductivity of composites by influencing the interface area and the formation of conductive networks~\cite{Jasmee2021,Ahmad2020}. Kim et al.~\cite{kim2016thermal} investigated the effect of graphene nanoplatelets (GNPs) in polymer composites and found that GNPs with larger lateral dimensions and minimal thickness exhibit higher thermal conductivities. The larger surface area of the GNPs enhances the interface between the filler and the matrix, facilitating better heat transfer across the interface and throughout the composite material, see Figure~\ref{fig:Kim}.
High aspect ratio fillers, such as carbon nanotubes (CNTs) and nanofibers, can form percolation networks within the matrix, providing continuous pathways for heat flow~\cite{Dong2021,Fang2021}. The formation of such networks reduces thermal resistance and significantly enhances the overall conductivity of the composite. The effectiveness of these networks depends on the ability to achieve uniform dispersion and proper orientation of the nanofillers within the matrix~\cite{Zhan2020,Guo2021}.

\subsubsection{Impact of spatial distribution and dispersion}
The spatial distribution and dispersion of reinforcements within the matrix significantly influence heat conduction~\cite{Wu2021,Ren2021}. Uniform dispersion of fillers leads to isotropic thermal properties, where the thermal conductivity is similar in all directions. In contrast, clustering or agglomeration of fillers can introduce anisotropy and create localized regions with varying thermal conductivities~\cite{kayhani2009exact}.
Clustering impedes the formation of effective heat conduction pathways and introduces additional interfaces that scatter phonons, thereby reducing thermal conductivity. Achieving a homogeneous distribution of reinforcements is crucial for optimizing the thermal performance of composites~\cite{zhang2020superior}. Techniques such as surface functionalization of fillers and advanced mixing methods are employed to enhance dispersion and prevent agglomeration~\cite{Wu2021,wang2024graphene,Li2021}.

\subsubsection{Impact of volume fraction of reinforcements}
The volume fraction of the reinforcement phase is another critical factor influencing the thermal conductivity of composites~\cite{pasztory2021overview}. Increasing the amount of high-conductivity filler generally enhances the composite's overall thermal conductivity due to the increased number of conductive pathways. However, there is often a percolation threshold, above which the fillers form a continuous network that dramatically improves thermal conduction~\cite{Kargar2018}.
In polymer composites filled with carbon-based materials like CNTs or graphene, the thermal conductivity increases with the filler content up to a certain point. Beyond this point, further additions may lead to diminishing returns or even adverse effects on mechanical properties and processability~\cite{guo2020factors}. Therefore, optimizing the filler content is necessary to balance thermal performance with other material properties.

\subsection{Convective Heat Transfer in Composite Materials}
Convective heat transfer in composite materials involves the movement of heat through the motion of fluids either within the material (internal convection) or over its surface (external convection)~\cite{nield2006convection}. While composites are primarily solid materials, the incorporation of design features such as porosity, internal channels, and specific surface textures can significantly influence convective heat transfer processes~\cite{xu2019review}. Understanding and optimizing these features are crucial for applications where efficient heat dissipation or absorption is required.

\subsubsection{Internal Convection in Porous Composites}
Internal convection occurs when a fluid moves within the voids or channels of a porous material, enhancing heat transfer through the material~\cite{habibishandiz2022critical}. In porous composites, pore sizes typically range from a few nanometers to several micrometers, which can significantly influence the effectiveness of diffusion processes~\cite{prokevsova2006porosity,lawrence2017porosity}. In these materials, internal convection usually occurs when pore sizes exceed one millimeter~\cite{apostolopoulou2021thermally}; see Figure~\ref{fig:Convection}. Therefore, enhancing convection efficiency can be achieved by integrating designed porosity or internal channels that facilitate fluid flow~\cite{nield2006convection}. This approach is particularly beneficial in applications such as heat exchangers, thermal insulation systems, and cooling structures where efficient heat transfer is critical~\cite{dai2024strategic}.
Deliberately tailored porosity, including the precise control of pore size, shape, and distribution, enables optimization of heat transfer rates and thermal management efficiency in these systems~\cite{lei2024characteristics}.
Alhashash et al.~\cite{alhashash2023free} studied the free convective heat transfer in porous composite materials and demonstrated that the strategic placement of voids or channels can significantly enhance convective heat transfer. By optimizing the size, shape, and distribution of these voids or channels, it is possible to control the fluid flow pathways within the composite, thereby increasing the convective heat transfer coefficient~\cite{satyamurty2010forced}.
The enhancement of internal convection in composites can be achieved through various design strategies.
For example, embedding micro-channels within the composite allows for the circulation of cooling or heating fluids~\cite{naqiuddin2018overview,xu2024novel}. The increased surface area provided by these channels facilitates greater heat exchange between the fluid and the composite material~\cite{huang2024adoption}.
In addition, selecting matrix materials with higher permeability enables better fluid flow through the composite, enhancing convective heat transfer. This is particularly important in applications where natural convection is utilized. Designing composites with graded porosity can optimize fluid flow and heat transfer. Higher porosity regions allow for increased fluid movement, while lower porosity areas provide structural support~\cite{wang2021three}.
Boyard~\cite{boyard2016heat} emphasized that composites designed with internal porosity or micro-channels allow for better fluid passage, enhancing convective heat transfer by increasing the contact surface area between the fluid and the solid matrix. This design not only improves heat transfer rates but can also contribute to weight reduction, which is beneficial in applications such as aerospace and automotive industries~\cite{singh2018review}.

\subsubsection{External Convection and Surface Characteristics}
External convective heat transfer refers to the transfer of heat between the surface of a composite material and the surrounding fluid, such as air or liquid coolant~\cite{xu2019review}. The efficiency of this process is influenced by the surface texture, roughness, and overall geometry of the composite material.
The surface characteristics of composites can be engineered to enhance convective heat transfer by modifying surface roughness, or adding surface features, or optimizing surface geometry~\cite{nguyen2021comprehensive}.
Aghababaei et al.~\cite{aghababaei2020determination} found that the surface texture and finish of composites significantly influence external convective heat transfer by altering fluid flow patterns near the surface. Increased surface roughness can disrupt the laminar boundary layer, promoting turbulence and enhancing heat transfer rates.
Incorporating fins, ribs, or other protrusions on the surface increases the effective surface area and can improve heat dissipation~\cite{lan2024study}. These features can be integrated into the composite design during manufacturing.
The overall shape and orientation of the composite component affect the flow of fluid over its surface. Streamlined shapes can enhance convective cooling by promoting favorable flow patterns.
The impact of surface characteristics on convective heat transfer is crucial in applications such as electronic cooling systems, where composite materials are used as heat sinks or enclosures. By tailoring the surface properties, engineers can optimize the thermal performance of these components without significantly altering their size or weight~\cite{panda2021review}.

\subsubsection{Combined Conduction and Convection Effects}
\begin{figure}
    \centering
    \includegraphics[width=1\linewidth]{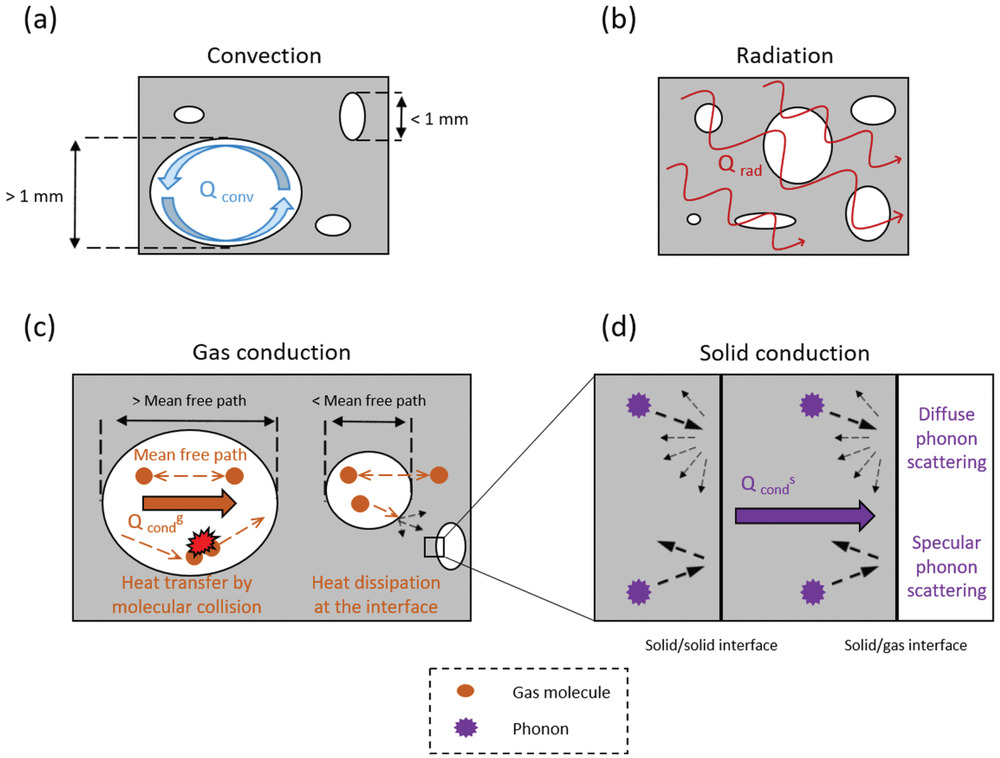}
    \caption{Illustration of the modes of heat transport in porous materials: (a) convection, (b) radiation, (c) gas conduction, including coupling effects at the gas–solid interface, and (d) solid conduction, highlighting diffuse and specular phonon scattering at interfaces~\cite{apostolopoulou2021thermally}. While conduction and radiation occur across multiple scales, from micro to macro, convection in composites typically manifests as a macroscopic phenomenon, requiring pore sizes larger than a millimeter.}
    \label{fig:Convection}
\end{figure}

In many composite materials, heat transfer involves a combination of conduction through the solid matrix and convection within internal voids or over the external surface. The interplay between these mechanisms can be complex and requires careful consideration in the design process.
The presence of internal convection can alter the effective thermal conductivity of a composite material~\cite{apostolopoulou2021thermally}. Fluid movement within the pores can enhance heat transfer beyond what would be expected from conduction alone; see Figure\ref{fig:Convection}.
In addition, in porous composites, thermal dispersion caused by fluid flow can contribute to heat transfer~\cite{kasaeian2017nanofluid}. This effect is analogous to dispersion in mass transfer and can enhance the overall heat transfer rate.
Moreover, the directional nature of internal channels or porosity can lead to anisotropic thermal properties, where heat transfer rates vary depending on the direction~\cite{nakajima2007fabrication}. This can be advantageous in applications requiring directional heat dissipation~\cite{li2024advances}.
Kayhani et al.~\cite{kayhani2009exact} provided an exact solution for heat conduction in particulate composites with thermal contact resistance, highlighting the importance of considering both conduction and convection in the thermal analysis of composites. Understanding the combined effects allows for more accurate predictions of thermal behavior and better design of composite materials for specific applications.

\subsection{Radiative Heat Transfer in Composites}
Radiative heat transfer becomes significantly important at high temperatures, especially in composite materials that contain semi-transparent constituents or are exposed to environments where radiation dominates the heat transfer process~\cite{retailleau2020experimental,howell2020thermal}. Unlike conduction and convection, radiative heat transfer involves the emission, absorption, and transmission of electromagnetic radiation, primarily in the infrared spectrum for thermal applications~\cite{zhu2023electrical}.

In composite materials, the radiative properties are not solely a function of the individual components but are also heavily influenced by the microstructure, arrangement of constituents, surface characteristics, and interaction with the surrounding environment~\cite{hemath2020comprehensive}. Understanding and controlling these factors are crucial for optimizing thermal performance in applications ranging from high-temperature insulation to thermal management systems in aerospace and electronics.
Wei et al.~\cite{wei2013radiative} studied the radiative properties of silica aerogel composites, highlighting their effectiveness as thermal insulators in high-heat environments due to their high emissivity and low density. Silica aerogels are known for their porous structure, which significantly reduces heat transfer by conduction and convection. However, their semi-transparent nature allows for radiative heat transfer, which can be substantial at elevated temperatures. By incorporating opacifiers or adjusting the pore size, the radiative heat transfer can be minimized, enhancing the overall insulating performance of the aerogel composites.
Surface emissivity plays a critical role in the radiative heat transfer of composites. The emissivity of a material determines its ability to emit or absorb thermal radiation compared to a perfect blackbody~\cite{karwa2020laws}. High-emissivity materials can efficiently radiate heat away, which is advantageous in applications requiring rapid heat dissipation~\cite{he2009high}. Conversely, low-emissivity materials reflect thermal radiation, making them suitable for thermal insulation purposes~\cite{jelle2015low,akamine2024radiation}.

Gasilov et al.~\cite{gasilov2023calculation} investigated the use of Z-pinch composites in high-energy environments, demonstrating how materials with high emissivity can efficiently manage heat through radiation. In such extreme conditions, conductive and convective heat transfer mechanisms may be insufficient or impractical for thermal management. By optimizing the emissivity of the composite surface, it is possible to enhance radiative cooling, thus protecting the structural integrity of the material under intense thermal loads~\cite{zhuang2023preparation,saad2023radiation}.
The environmental interaction significantly influences the effectiveness of radiative heat transfer in composites~\cite{howell2020thermal}. Factors such as surrounding temperature, presence of other radiative bodies, and the nature of the ambient medium affect the net radiative heat exchange~\cite{cuevas2018radiative}. In high-temperature applications, the temperature difference between the composite surface and its environment drives the radiative heat transfer. For instance, in vacuum or space environments where convection is negligible, radiation becomes the dominant mode of heat transfer~\cite{sidebotham2015heat}.

The spectral properties of composite materials also impact radiative heat transfer~\cite{rousseau2016thermal}. The emissivity and absorptivity can vary with wavelength, especially in materials with complex microstructures or multi-phase constituents. By engineering the spectral emissivity, composites can be tailored to emit or absorb radiation efficiently at specific wavelengths~\cite{retailleau2020experimental}. This is particularly important in applications like thermal control coatings, infrared stealth technology, and thermophotovoltaic systems~\cite{gamel2021review}.
Microstructural features such as porosity, grain boundaries, and interfaces influence the radiative properties of composites. In materials where the constituents have different refractive indices, scattering of radiation occurs at the interfaces, affecting the overall radiative heat transfer~\cite{yang2015comparative}. Techniques to model and predict radiative transfer in such heterogeneous media involve complex calculations, often requiring numerical methods like the Monte Carlo ray-tracing technique or the discrete ordinates method~\cite{howell2021past}.
Furthermore, the inclusion of radiatively active additives or coatings can modify the radiative behavior of composites~\cite{suryawanshi2009radiative}. For example, adding carbon black or metal oxide particles to a composite can increase absorption and scattering, thereby reducing radiative transmission and enhancing insulation performance~\cite{meng2009absorption}. Conversely, applying reflective coatings or incorporating metallic fibers can decrease emissivity, reflecting thermal radiation and reducing heat gain~\cite{wijewardane2012review}.

\section{Modeling and Predicting Heat Transfer in Composite Materials} \label{section:modeling}
\begin{figure}
    \centering
    \includegraphics[width=0.6\linewidth]{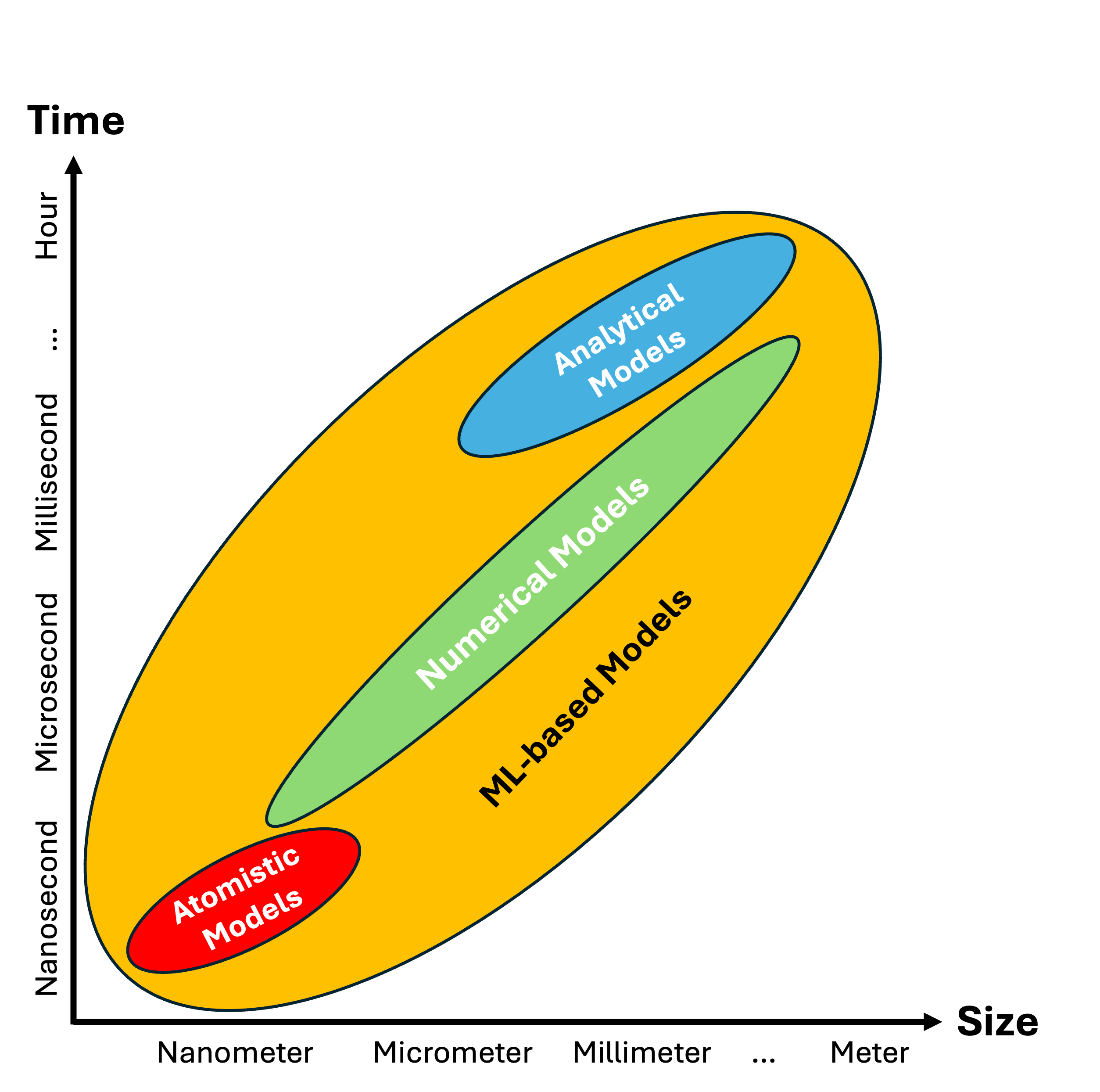}
    \caption{Illustration of the applicability of different modeling approaches for predicting heat transfer in composite materials across diverse size and time scales. Each method demonstrates unique advantages tailored to the specific scale and complexity of the problem being analyzed.}
    \label{fig:scale}
\end{figure}

Predicting heat transfer in composite materials relies on a combination of analytical, numerical, and atomistic simulation methods, alongside machine learning (ML) techniques. These approaches enable the design and optimization of composites, allowing precise control over thermal properties critical for high-performance applications in industries such as aerospace, automotive, and electronics. Each method offers specific advantages and limitations, demonstrating varying efficiencies depending on the size and time scales being analyzed, as illustrated in Figure~\ref{fig:scale}.

\subsection{Analytical Models}
Analytical models serve as fundamental tools in predicting heat transfer within composite materials. These models provide simplified solutions that help to grasp the underlying physics of thermal processes, making them essential in the initial stages of composite design and analysis. Analytical models in the context of studying heat transfer in composite materials are generally applied at different scales, from micro to macro, depending on the nature of the model and the specific properties being investigated.
    \begin{itemize}
        \item \textbf{Rule of Mixtures:} This basic yet crucial model estimates the effective thermal conductivity by averaging the properties of the composite's constituents based on their volume fractions. While useful for preliminary design considerations, this approach tends to oversimplify the effects of microstructural features. This model fails to account for the complexities arising from the anisotropic nature of composites and the interface phenomena between different materials~\cite{burger2016review}.
        \item \textbf{Effective Medium Theories (EMT):} These theories enhance the Rule of Mixtures by incorporating the influence of the shape, size, and distribution of particles within the matrix~\cite{Taleb2022}. Effective medium theories provide a more nuanced prediction of thermal behavior by considering how these microstructural variables affect the composite's overall thermal conductivity. They show how different configurations can impact the path of heat flow and the effective thermal resistance of the composite material~\cite{lebeda2024shaping}.
        \item \textbf{Homogenization Techniques:} Especially useful for composites with periodic microstructures, homogenization techniques aim to determine effective macroscopic properties by averaging field variables over a representative volume element (RVE). This approach is particularly adept at handling materials with regular geometrical patterns, as demonstrated by Kayhani et al.~\cite{kayhani2009exact}, who applied these techniques to analyze heat transfer in cylindrical composite laminates. Their work provides an exact solution for steady-state conduction, highlighting how effective homogenization can be in predicting thermal distributions in structured composites~\cite{kayhani2009exact}.
    \end{itemize}
Despite their utility for quick assessments and theoretical understanding, the limitations of analytical models stem from the assumptions they require, such as material homogeneity and isotropy, which are often not present in real-world composites. These limitations necessitate subsequent validation and refinement through more detailed experimental or numerical methods. The analytical methods lay the groundwork for initial estimates but often require supplementation through empirical data or more sophisticated simulations to ensure accuracy and reliability in practical applications.

\subsection{Numerical Methods}
Numerical simulations are indispensable tools for analyzing the thermal behavior of composite materials. These methods offer the flexibility needed to address the complex geometries, non-uniform material distributions, and various boundary conditions typical of composites~\cite{zhao2020review,Wang2023}.
    \begin{itemize}
        \item \textbf{Finite Element Analysis (FEA):} FEA is a cornerstone technique for modeling heat transfer in composites, commonly applied across multiple scales ranging from the mesoscale to the macroscale. At the mesoscale, it evaluates the interactions and thermal behavior of composite structures on the level of material clusters or phase interactions, typically spanning from micrometers to millimeters~\cite{Ghobadian2021}. At the macroscale, FEA is employed to assess the overall thermal performance of larger composite components or systems, which can range from several millimeters to meters, providing crucial insights into their behavior under real-world operational conditions. It works by discretizing complex geometries into smaller, manageable elements, each with its own set of properties, enabling precise numerical solutions to heat transfer equations~\cite{sharma2022finite}. This method can accommodate variations in material properties across different components of the composite, as well as complex boundary conditions involving thermal and mechanical loads~\cite{jiang2017numerical,sun2024multi}.
        \item \textbf{Computational Fluid Dynamics (CFD):} While primarily recognized for its application in fluid flow studies, CFD is also crucial for modeling convective heat transfer within and around composite structures~\cite{cruz2022computational}. At the mesoscale, CFD effectively explores the interactions between fluid flows and the microstructures of composites, such as pores or channels within the material, typically spanning from micrometers to millimeters. This method extends to the macroscale, where it is particularly effective for analyzing the dynamic interactions between fluid flows and the larger structural boundaries of composite systems, which can range up to several meters. Alhashash utilized CFD to investigate the behavior of nanofluids within composite enclosures, illustrating how this approach can be employed to optimize thermal management strategies in engineering applications~\cite{alhashash2023free}.
    \end{itemize}
Both FEA and CFD are enhanced by advances in computing power, which allow for more detailed simulations over larger scales and more complex conditions. These methods are not just tools for prediction but also serve as a means of verification for theoretical models, ensuring that designs are both efficient and practical before physical prototypes are developed.
Furthermore, the integration of FEA and CFD can provide a comprehensive understanding of the thermal and mechanical performance of composites. This synergy is especially important in applications such as aerospace and automotive industries, where both heat transfer and structural integrity are critical~\cite{anwar2019computational}. For instance, a combination of these methods were utilized to analyze the thermal anisotropy in carbon fiber reinforced composites, providing insights into how directional properties can be exploited to guide heat flow effectively within a material~\cite{lebeda2024shaping}.

\subsection{Atomistic Simulation Methods}
Atomistic simulations, such as Molecular Dynamics (MD)~\cite{hollingsworth2018molecular}, offer valuable insights into the intricate mechanisms of heat transfer that are not observable at the macroscopic level. As shown in Figure~\ref{fig:scale}, atomistic simulation methods are currently constrained to dimensions less than a micrometer and systems of fewer than a million atoms due to computational limitations. However, these methods have the potential to be applied at larger scales if computational power, such as that offered by quantum computing~\cite{Sood2024,DeLeon2021,Thi2024}, advances significantly, achieving 100 to 1000 times the current computational capacity. 

\begin{figure}
    \centering
    \includegraphics[width=0.6\linewidth]{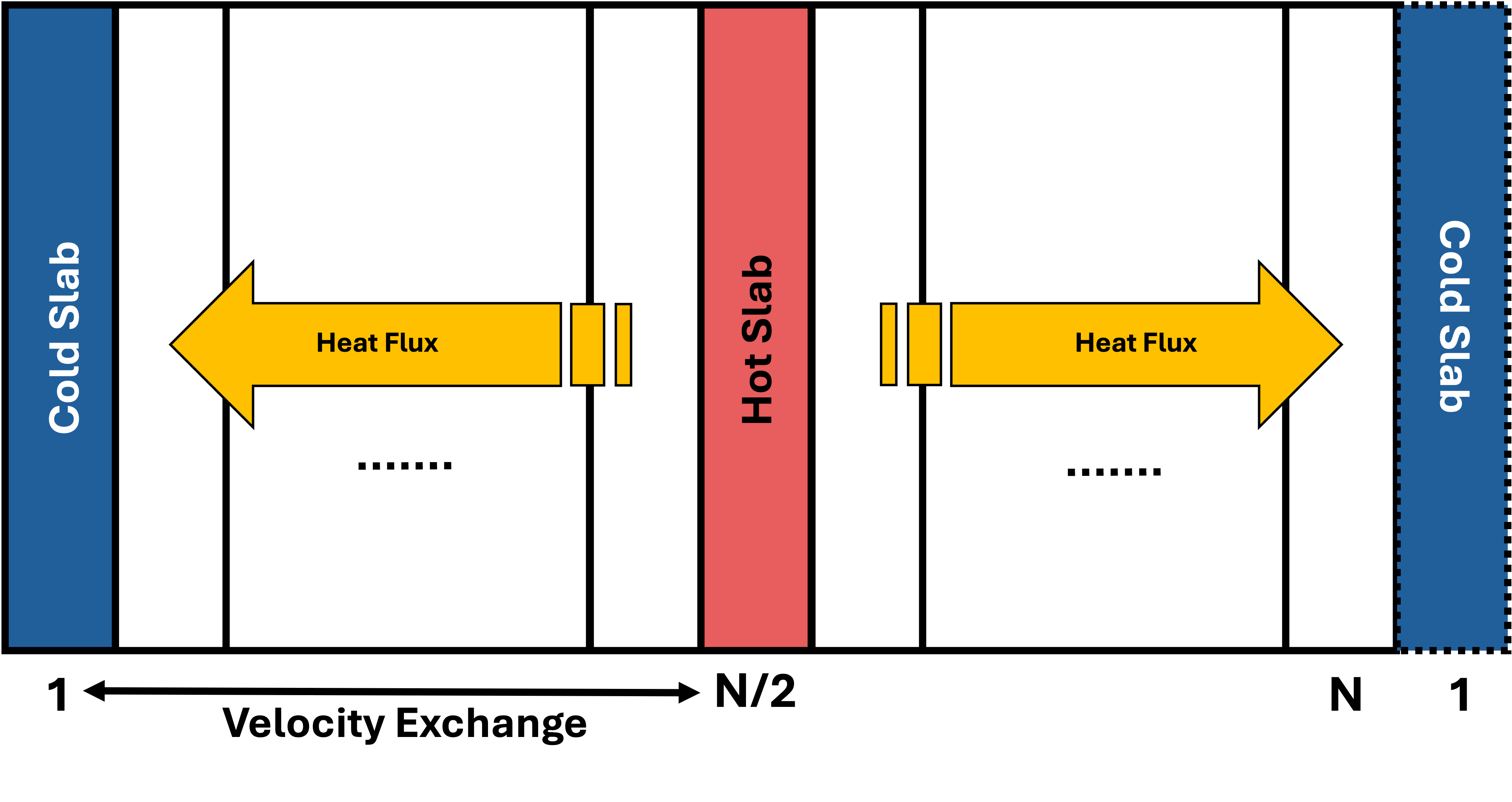}
    \caption{Schematic illustration of the nonequilibrium molecular dynamics (NEMD) method for calculating thermal conductivity. In this approach, the heat flux (effect) is imposed on the system, while the resulting temperature gradient (cause) is obtained from the simulation, effectively reversing the traditional cause-and-effect relationship. This method simplifies implementation and offers advantages such as compatibility with periodic boundary conditions, conservation of total energy and linear momentum, and efficient sampling of the rapidly converging temperature gradient instead of the slowly converging heat flux.}
    \label{fig:RNEMD}
\end{figure}

MD methods are especially crucial for understanding and predicting the thermal properties of composites where classical theories might fall short~\cite{dong2018equivalence,najmi2023review}.
MD simulations serve as a powerful tool to explore phonon interactions, interface scattering, and the effects of nanoscale defects on thermal transport~\cite{McGaughey2006,Rahman2021}. These simulations are particularly useful for materials requiring precise thermal management strategies, such as those in energy storage and electronic devices~\cite{fang2023thermal}.
Within MD, methods such as the Green-Kubo method~\cite{Knoop2023} and Reverse Non-Equilibrium Molecular Dynamics~\cite{muller1997simple,muller1999cause,muller2004reverse} often used for calculating the thermal conductivity.
    \begin{itemize}
        \item \textbf{Green-Kubo Method:} The Green-Kubo method is a widely used approach for calculating thermal conductivity based on the autocorrelation function of the heat current in an equilibrium state~\cite{Knoop2023}. This method relies on the fluctuation-dissipation theorem, which connects microscopic energy fluctuations to macroscopic transport properties~\cite{ford2017fluctuation}. It is particularly effective for materials with homogeneous or periodic microstructures, such as crystalline solids or well-ordered composite systems. The Green-Kubo method provides a direct and reliable measure of thermal conductivity, making it an essential tool for understanding how microscopic properties, such as atomic vibrations and phonon transport, influence macroscopic thermal behaviors~\cite{kang2017first,manjunatha2021atomic}. Additionally, this approach is instrumental in validating theoretical predictions and serves as a benchmark for other computational techniques, particularly in studies involving molecular dynamics simulations~\cite{knoop2023ab}. Despite its strengths, the method can be computationally expensive for large or highly disordered systems, requiring careful optimization of simulation parameters~\cite{dongre2017comparison}.
        \item \textbf{Reverse Non-Equilibrium Molecular Dynamics (RNEMD):} RNEMD induces a heat flux by artificially creating a temperature gradient within the simulation box, allowing for the direct measurement of thermal conductivity~\cite{muller1997simple,muller1999cause,muller2004reverse}, see Figure~\ref{fig:RNEMD}. This method bypasses the need for equilibrium simulations, making it particularly suitable for studying materials with complex microstructures or non-linear thermal behavior~\cite{degirmenci2020reverse,gulzar2019energy,zanane2024thermal,shavalier2024reverse}. RNEMD is instrumental in exploring how structural or compositional changes influence thermal properties, such as phonon scattering and heat flow pathways~\cite{breitkopf2024theoretical}. Recent advancements have expanded its applicability, including the integration of RNEMD with machine-learned interatomic potentials to enhance the accuracy and efficiency of simulations~\cite{wang2022thermal,cao2024thermal,wan2024thermal,dong2024molecular}. For instance, this approach has been utilized to predict thermal transport properties in complex materials like $\alpha$, $\beta$, and $\epsilon$-Ga$_2$O$_3$, highlighting its capability for material optimization in applications requiring specific thermal functions~\cite{dong2024molecular}. Moreover, RNEMD provides critical insights into anisotropic thermal conductivity in layered and heterogeneous systems, where traditional methods may fall short~\cite{alaghemandi2011thermal,alaghemandi2012thermal}. Despite its strengths, RNEMD requires careful calibration of simulation parameters, such as the length of the simulation box and the imposed temperature gradient, to ensure accurate and reliable results. These capabilities make RNEMD a valuable tool in the design and development of advanced materials for high-performance thermal management~\cite{dongre2017comparison}.
    \end{itemize}
These atomistic techniques allow for a nuanced understanding of thermal processes at the atomic level, enabling the design of composite materials that meet specific performance criteria~\cite{Gu2021}. Using MD simulations alongside experimental data and theoretical models, researchers can enhance the accuracy of predictions and improve material designs, ensuring composites perform optimally in their respective applications.

\subsection{AI-based Models}
The integration of artificial intelligence (AI) and machine learning (ML) with traditional modeling techniques represents a transformative advancement in materials science~\cite{schmidt2019recent,ma2024mastery}. By leveraging large datasets, AI enables rapid and accurate prediction, optimization, and design of composite materials, significantly reducing the reliance on costly and time-intensive experimental trials~\cite{shen2023predicting,rong2019predicting}.

\begin{itemize}

\item \textbf{Predictive Modeling:} Machine learning algorithms excel in predictive modeling by analyzing extensive datasets to forecast the thermal and mechanical behaviors of composites. For instance, convolutional neural networks (CNNs) have been employed to establish microstructure-property relationships with exceptional accuracy~\cite{rong2019predicting,liu2022stochastic}. In a recent study, Shen et al.\cite{shen2023predicting} demonstrated the use of CNNs to predict the effective thermal conductivity of fibrous and particulate composites, outperforming traditional finite element and homogenization methods in both speed and accuracy. 
Similarly, Sang et al.\cite{sang2023accurate} developed random forest models to predict the microstructural configurations of binary composites using elastic wave data, achieving a prediction accuracy of over 95$\%$.

\item \textbf{Optimization Algorithms:} AI models are increasingly used for optimizing the configurations of composites to enhance desired properties. For example, Liu et al.\cite{liu2024single} utilized gradient boosting and random forest algorithms to predict and optimize the thermal conductivity of polymer composites with single fillers, while identifying key factors such as filler volume fraction and matrix conductivity. Similarly, Mukherjee et al.\cite{mukherjee2024predicting} applied supervised learning models to predict the thermal conductivity of hollow glass microsphere composites, showcasing the ability of ML to optimize material compositions pre-synthesis, thus reducing experimental trial and error.

\item \textbf{Multi-scale Integration:} One of the most promising applications of AI in composite modeling lies in bridging multi-scale interactions. Liu et al.~\cite{liu2022stochastic} employed a hierarchical multi-scale approach, integrating atomistic simulations at the nanoscale with continuum models at the macro level, to predict thermal conductivity in carbon nanotube-polymer composites. By coupling machine learning with multi-scale modeling, this approach provides a computationally efficient alternative to traditional methods.

\item \textbf{Explainable AI (XAI):} A significant challenge in AI is the lack of transparency and interpretability in how complex machine learning models, such as deep learning and neural networks, make decisions~\cite{linardatos2020explainable,shah2021neural,csahin2024unlocking}. These models often act as "\textit{black boxes}", where their internal processes are not easily understandable, even by experts. Addressing this issue involves developing AI systems that offer human-interpretable explanations for their predictions and decisions~\cite{li2022interpretable}. 
XAI has gained significant traction in materials science, especially for predicting thermal and thermophysical properties of composites~\cite{barua2024interpretable,cakiroglu2024explainable,qayyum2023explainable}. For instance, Huang et al.~\cite{huang2024interpretable} proposed an interpretable deep learning strategy for predicting the effective thermal conductivity of porous materials. This methodology not only improved prediction accuracy but also provided actionable insights for optimizing porous material designs~\cite{huang2024interpretable}.
Such developments underscore the potential of XAI to bridge the gap between complex AI models and practical applications in composite material design~\cite{mikulskis2019toward,dean2023interpretable}. By making AI models interpretable, XAI enables researchers and engineers to validate predictions, refine material structures, and enhance quality assurance processes~\cite{rudin2022interpretable,oviedo2022interpretable}. This interpretability is critical for ensuring the adoption of AI-driven solutions in high-stakes applications where accountability and accuracy are paramount~\cite{allen2022machine}.

\end{itemize}

\section{Experimental Techniques for Measuring Heat Transfer in Composite Materials}
Accurately measuring the heat transfer properties of composite materials is essential for evaluating their performance in applications such as aerospace, automotive, and electronics. Various experimental techniques are employed to determine thermal properties like thermal conductivity, diffusivity, and specific heat capacity~\cite{carlsson2014experimental}. The choice of technique depends on factors such as the material's properties, geometry, and the required measurement precision. Below is a list of commonly used experimental methods for measuring thermal properties in composite materials.

\begin{itemize} 
    \item \textbf{Steady-State Techniques}: 
        \begin{itemize} 
            \item \textbf{Guarded Hot Plate Method}: A widely used technique for determining the thermal conductivity of homogeneous and layered materials. The method involves placing the composite specimen between two plates—one heated and one maintained at a constant lower temperature. A guard ring minimizes lateral heat losses, ensuring one-dimensional heat flow through the specimen. Thermal conductivity is calculated once a steady-state temperature gradient is established. While highly accurate, this method can be time-consuming due to the need to reach thermal equilibrium~\cite{assaf2017measurement,naseem2019simple}.
            \item \textbf{Heat Flow Meter Method}: Similar to the guarded hot plate method, the specimen is sandwiched between a hot and a cold plate. However, it uses heat flux transducers to directly measure the heat flow through the specimen. This method offers faster measurements with reasonable accuracy, making it suitable for quality control and routine testing~\cite{elkholy2022accurate}.
            \item \textbf{Comparative Longitudinal Heat Flow Method}: Also known as the comparative cut-bar method, this technique places the test specimen between two reference materials with known thermal conductivities. By measuring the temperature gradient along the assembly, the thermal conductivity of the specimen is determined through comparison. Modifications like integrating insulated guards help minimize heat losses and improve measurement accuracy~\cite{elkholy2022accurate}.
    \end{itemize}
    \item \textbf{Transient Techniques}:
        \begin{itemize}
            \item \textbf{Laser Flash Analysis (LFA)}: This method measures the thermal diffusivity of a specimen by applying a short energy pulse, usually from a laser, to its front face and recording the temperature rise on the rear face. The time it takes for the heat to traverse the specimen is used to calculate thermal diffusivity. LFA is particularly effective for thin materials or those with high thermal diffusivity and allows for rapid measurements over a range of temperatures~\cite{elkholy2022accurate}.
            \item \textbf{Transient Plane Source (TPS) Method}: Also known as the hot disk method, TPS employs a planar sensor that acts as both a heat source and a temperature sensor. Placed between two pieces of the specimen or on its surface, the sensor records the temperature response over time. This method can measure both thermal conductivity and diffusivity and is versatile enough to handle anisotropic materials, making it suitable for complex composites~\cite{kearney2022measurement}.
            \item \textbf{Hot Wire Method}: In this technique, a thin wire embedded in the specimen serves as a heat source when an electrical current passes through it. The temperature rise is measured over time, and the thermal conductivity is calculated based on the rate of temperature increase. The hot wire method is suitable for fluids and some solid materials and offers rapid measurements~\cite{naseem2019simple}.
        \end{itemize}
    \item \textbf{Infrared Thermography}: A non-destructive technique that involves heating the specimen and using infrared cameras to monitor the surface temperature distribution in real-time~\cite{abad2017non}. Infrared thermography is effective for identifying defects, inhomogeneities, and evaluating both surface and subsurface thermal properties of composite materials~\cite{kearney2022measurement}. Changes in color at different temperatures allow for qualitative and quantitative assessment of heat flow patterns, making it particularly useful for inspecting large areas quickly.
    \item \textbf{Differential Scanning Calorimetry (DSC)}: DSC measures the heat flow into or out of a material as a function of temperature or time. It is used to determine specific heat capacity, phase transitions, and other thermal properties. While primarily employed for studying thermal transitions, DSC provides valuable data on the heat capacity of composite materials, which is essential for comprehensive thermal analysis~\cite{assaf2017measurement}.
\end{itemize}

Each technique offers unique advantages and is selected based on the specific requirements of the material system and the properties to be measured. Steady-state methods like the guarded hot plate provide high accuracy for thermal conductivity measurements but require longer times to reach equilibrium. Transient methods such as LFA and TPS offer rapid measurements and can accommodate materials with varying thermal properties. Non-destructive techniques like infrared thermography are valuable for assessing internal structures and detecting defects without damaging the specimen~\cite{palacios2019thermal}. The integration of multiple techniques often provides a more comprehensive understanding of the thermal behavior of composite materials.

\section{Applications of Composite Materials in Heat Transfer Management}
\begin{figure}
    \centering
    \includegraphics[width=1\linewidth]{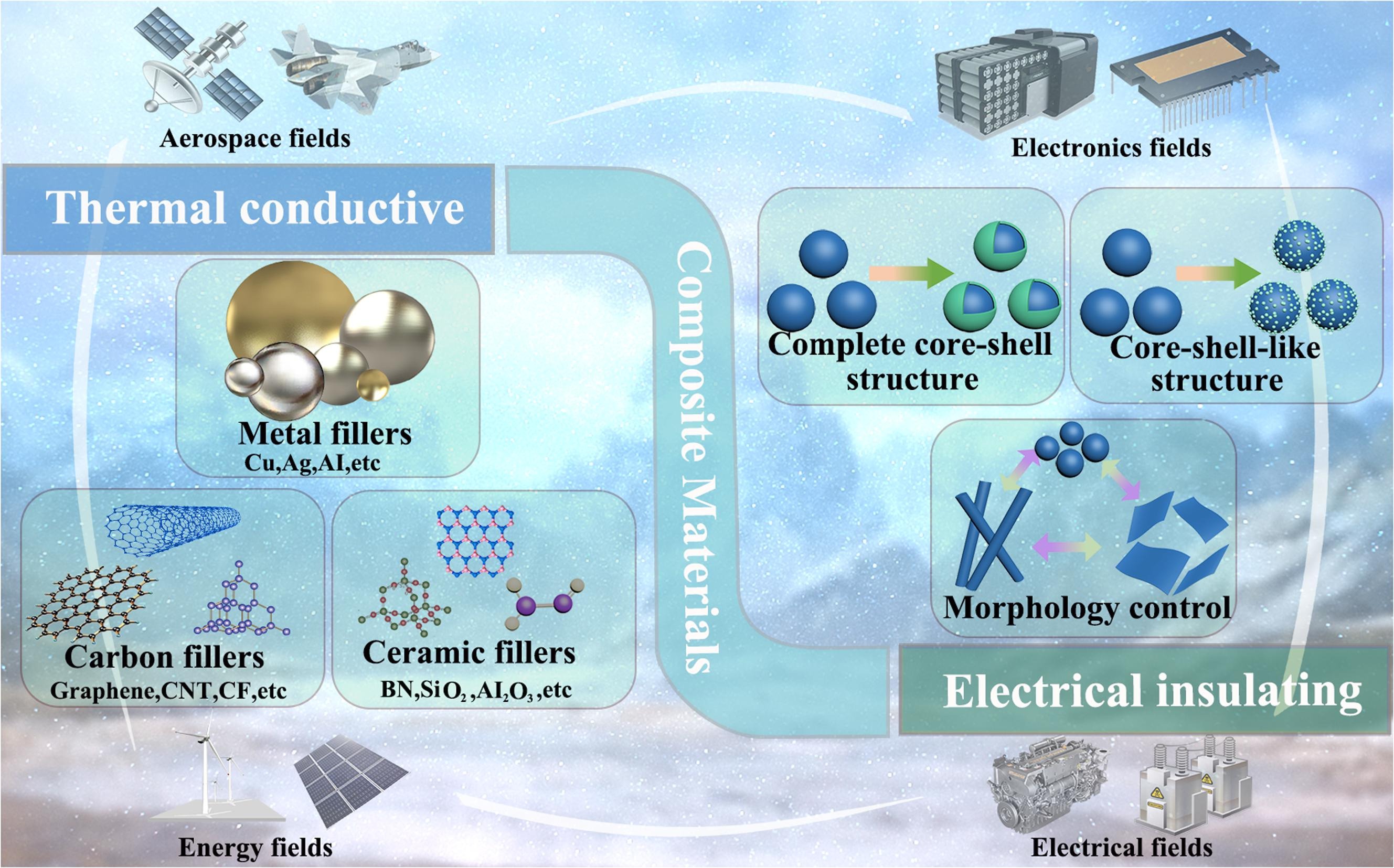}
    \caption{Composite materials engineered to balance high thermal conductivity with high electrical insulation properties, achieving desired performance characteristics in thermal management applications~\cite{chen2023recent}.}
    \label{fig:application}
\end{figure}

Composite materials have become indispensable in various industries due to their unique combination of mechanical strength, lightweight properties, and customizable thermal characteristics~\cite{bhong2023review}. In heat transfer and thermal management applications, composites offer significant advantages over traditional materials, enabling improved performance, efficiency, and reliability~\cite{kavimani2025evolution}. In many applications, composite materials are engineered to achieve a balance between thermal conductivity and electrical conductivity, as shown in Figure~\ref{fig:application}, to meet specific property requirements~\cite{chen2023recent}. This section explores the diverse applications of composite materials in heat transfer across multiple sectors, including aerospace, automotive, electronics, renewable energy, and construction.

\subsection{Aerospace Industry}
In the aerospace sector, thermal management is critical due to the extreme temperatures and rapid thermal cycling experienced during operation~\cite{van2022aircraft}. Composite materials are employed in several key areas. For instance, in thermal protection systems (TPS) for spacecraft and high-speed aircraft, advanced composites such as carbon-carbon composites and ceramic matrix composites (CMCs) provide thermal insulation and maintain structural integrity at high temperatures~\cite{meola2018composite,uyanna2020thermal,sehgal2022comparative}. These materials can withstand temperatures exceeding 2000°C, which is essential for re-entry vehicles and hypersonic aircraft.
Moreover, composites are used in engine components like turbine blades, nozzles, and combustion chambers, where materials must endure high temperatures and thermal gradients~\cite{nobrega2024review}. CMCs offer high-temperature stability, oxidation resistance, and reduced weight compared to metal alloys, enhancing engine efficiency and longevity~\cite{wang2021advances}. Additionally, lightweight composite heat exchangers and radiators improve thermal efficiency while reducing the overall weight of the aircraft, contributing to fuel savings and increased payload capacity~\cite{nobrega2024review}. Their corrosion resistance also extends the service life of these components~\cite{careri2023additive}.

\subsection{Automotive Industry}
\begin{figure}
    \centering
    \includegraphics[width=1.0\linewidth]{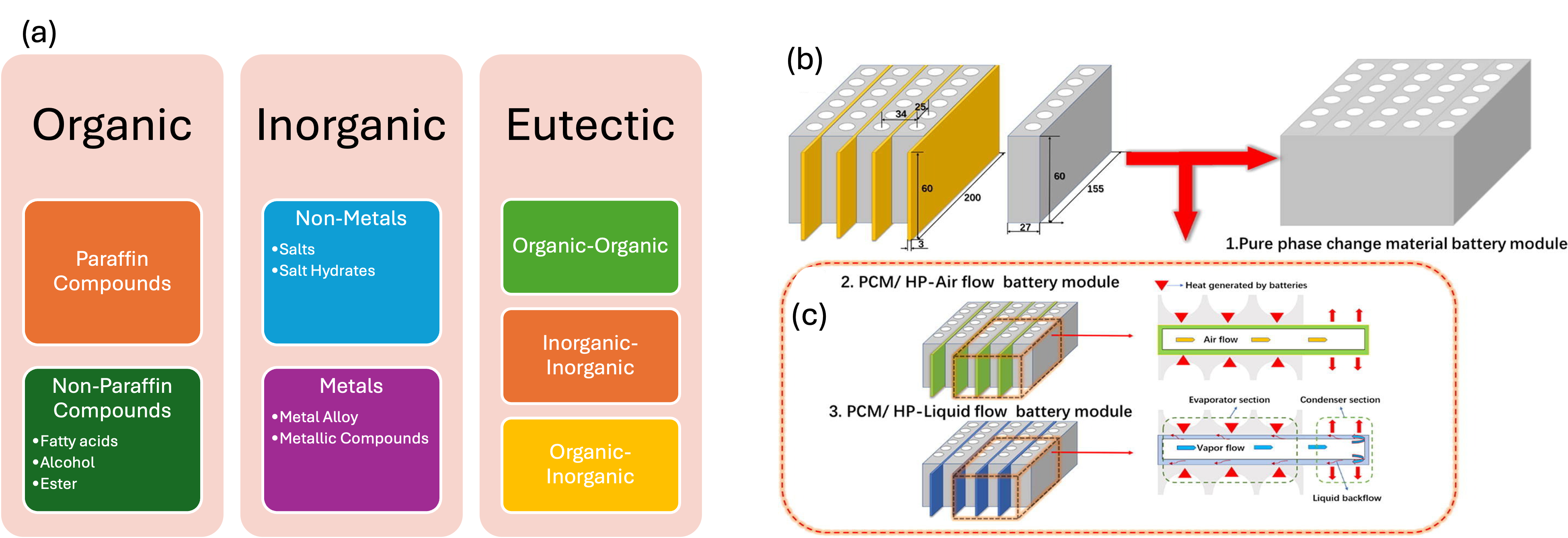}
    \caption{(a) Classification of various classes and types of phase change materials. (b) Schematic illustrations of the designed battery module and its sub-modules. (c) Representations of the PCM/Heat Pipe (HP) cooling systems~\cite{huang2018experimental}.}
    \label{fig:PCM}
\end{figure}

The automotive sector leverages composite materials for enhanced thermal management to improve performance and meet stringent emission regulations. In brake systems, carbon-fiber-reinforced composites (CFRCs) are used in high-performance brake discs and pads due to their excellent heat dissipation properties~\cite{khan2024advances}. They reduce brake fade under high-temperature conditions, improving safety and reducing wear.
Effective thermal management of batteries and power electronics is essential for electric vehicle (EV) performance and safety. 
An emerging approach in EV battery thermal management is the incorporation of phase change materials (PCMs) into composite structures~\cite{sanker2022phase}. PCMs have different classes and types as depicted in Figure~\ref{fig:PCM}(a). PCMs absorb excess heat generated during battery operation by undergoing a phase transition, typically from solid to liquid, thereby maintaining the battery temperature within optimal limits~\cite{huang2018experimental}. By integrating PCMs with thermally conductive composites in battery modules, it is possible to achieve passive thermal regulation without the need for active cooling systems, see Figure~\ref{fig:PCM}(b) and (c). This integration enhances energy efficiency and extends battery lifespan by preventing thermal runaway and improving overall safety~\cite{sanker2022phase,zhao2022comprehensive}.
Composites with high thermal conductivity also are used in battery casings and cooling systems to dissipate heat efficiently, enhancing battery life and reliability~\cite{togun2024critical}. Additionally, composite materials are employed in engine components such as intake manifolds and engine covers, providing thermal insulation to improve engine efficiency by reducing heat loss and minimizing thermal stress on surrounding parts.

\subsection{Electronics and Thermal Management}
With the miniaturization of electronic devices and increased power densities, thermal management has become a critical design consideration~\cite{yin2010thermal}. Composites are used in heat sinks and thermal interface materials (TIMs), where metal matrix composites (MMCs) and polymer matrix composites (PMCs) enhanced with thermally conductive fillers like graphene or carbon nanotubes facilitate efficient heat transfer from electronic components~\cite{bahru2021review,Wang2012}.
In printed circuit boards (PCBs), thermally conductive composite substrates are utilized to manage heat dissipation, improving the reliability and lifespan of electronic devices by preventing overheating~\cite{Wang2020}. 
Furthermore, composites with both electrical conductivity and thermal management capabilities enable the development of flexible and wearable electronics~\cite{Zhang2022-2}, where heat dissipation is crucial for performance and user safety~\cite{Li2022,Liu2024}.

\subsection{Renewable Energy Applications}
Composite materials contribute to the efficiency and durability of renewable energy systems. In wind turbine blades, glass-fiber and carbon-fiber composites provide the necessary strength-to-weight ratio and fatigue resistance~\cite{Brondsted2005}. The thermal stability of composites ensures performance under varying environmental temperatures, extending the operational life of the turbines~\cite{clarke2012thermal}.
In solar thermal collectors, composites with high thermal conductivity are used in absorber plates and piping systems to improve heat collection and transfer efficiency, enhancing overall energy capture~\cite{Haillot2017,Pradhan2020}. Additionally, as mentioned above, composites incorporating phase change materials enable efficient thermal energy storage, which is critical for balancing supply and demand in renewable energy systems~\cite{huang2018experimental}. These materials absorb and release heat during phase transitions, providing a stable energy supply.
Innovative applications of composites in thermal management include thermal barrier coatings (TBCs), which are applied to turbine blades and other high-temperature components~\cite{Mondal2021}. TBCs made from ceramic composites protect underlying materials from extreme heat, improving efficiency and extending service life~\cite{Zhang2016}.

\subsection{Building and Construction}
In the construction industry, composite materials enhance energy efficiency and safety~\cite{Hall2010}. Composite insulation panels provide superior thermal resistance, reducing energy consumption for heating and cooling~\cite{Zhang2022,Bhuiyan2023}. 
Materials like structural insulated panels (SIPs) combine insulation with structural support, streamlining construction processes~\cite{AbuJdayil2019}.
Composites are also used in heating, ventilation, and air conditioning (HVAC) components to improve thermal performance and reduce corrosion, enhancing system efficiency and lifespan~\cite{Choy2009}. Ducts made from composites are lighter and easier to install compared to traditional materials~\cite{Asim2022}. Fire-resistant composites are employed in structural components and barriers, providing thermal insulation and slowing the spread of fire, thus enhancing safety standards in buildings~\cite{Jefferson2024}.
\section{Recent Advances and Innovations in Heat Transport in Composite Materials}
The field of heat transport in composite materials is critical for applications in thermal management systems, energy storage, and thermal insulation. This research area has seen significant progress through experimental and theoretical approaches that aim to improve thermal conductivity, optimize thermal resistance, and design efficient thermal pathways within composites.

\subsection{Material Innovations and Design}
The design and development of novel materials remain central to enhancing thermal properties in composites. Innovations in this area have primarily focused on the integration of conductive fillers like graphene, carbon nanotubes, boron nitride, and metal particles into polymer matrices~\cite{Huang2016,Li2019,Zaaba2022,Salunke2021}. These fillers are chosen for their exceptional thermal properties and their ability to form percolative networks within the matrix, significantly improving the composite's overall thermal conductivity~\cite{Rajak2019,Kavimani2025}. Recent advances in synthesis and processing technologies have led to enhanced dispersion and precise alignment of these fillers, thus forming continuous and efficient thermal pathways that are critical for high-performance applications~\cite{Li2019, Rajak2019}.

Self-adaptive materials that dynamically adjust their thermal properties in response to environmental stimuli, such as temperature changes or mechanical stress, are becoming increasingly relevant~\cite{guo2022real,mei2022self,zhang2022thermal,sun2024application}. These materials can switch their thermal conductivity on demand, offering versatile solutions for smart thermal management systems in applications ranging from adaptive electronics to energy-efficient buildings~\cite{jaid2024metal,cui2019review}. For example, thermally responsive polymers that exhibit changes in their molecular structure when heated have been developed, allowing reversible adjustment of the material's thermal conductivity~\cite{Chen2023}.
Phase-change materials have been increasingly utilized to enhance thermal performance in composites~\cite{fan2011thermal,shamberger2020review,jaid2024metal}. By absorbing or releasing latent heat during phase transitions, PCMs can regulate temperature effectively, making them particularly useful in thermal management systems for electronics cooling and wearable devices. Recent advancements in microencapsulation techniques have improved the stability and efficiency of PCMs, preventing leakage and enhancing their thermal cycling reliability~\cite{sanker2022phase,das2020novel,el2024advanced}.

The development of bio-based polymers and the use of natural fillers offer alternatives to traditional composites, with the potential for reduced environmental impact~\cite{abdur2023review,hu2020nanocellulose,hassan2020acoustic}. These materials contribute to the circular economy and address growing concerns over resource depletion and material sustainability~\cite{azka2024review,alaghemandi2024sustainable}. These material innovations not only improve the fundamental thermal properties of composites but also pave the way for new applications and technologies that require sophisticated thermal management solutions.

\subsection{Advancements in Nanostructuring, Interfacial Engineering, and Additive Manufacturing}
Nanostructuring and interfacial engineering play pivotal roles in optimizing heat flow within composites, with additive manufacturing enhancing these capabilities through precise control over material structure and composition~\cite{parandoush2017review,yuan2021additive}. These integrated approaches are reshaping the design and fabrication of advanced thermal management systems~\cite{ghahfarokhi2021opportunities,freeman2023advanced,sun2024advanced}.

Techniques such as electrospinning and layer-by-layer assembly allow for meticulous control of the nanostructure within composites, optimizing thermal pathways and minimizing resistance~\cite{zhang2019progress}. This is exemplified in the use of vertically aligned carbon nanotube arrays and layered structures with graphene or boron nitride nanosheets, which create anisotropic composites exhibiting high directional thermal conductivity, beneficial for applications like heat spreaders in electronics~\cite{ivanov2006fast, li2017anisotropic, ji2018thermal}. Furthermore, hierarchical structuring, which combines fillers of varying sizes and shapes, forms multi-scale networks that enhance phonon transport and heat flow across different length scales~\cite{karger2020all, moreira2021review}.
Interfacial engineering enhances the effectiveness of these nanostructures by reducing thermal boundary resistance, crucial for maintaining efficient heat transport~\cite{liu2025engineering}. Techniques such as chemical functionalization and the use of coupling agents improve the interaction between fillers and the matrix~\cite{li2018surface}. Surface modifications, like grafting functional groups to fillers such as carbon nanotubes or graphene, boost their compatibility with polymer matrices, facilitating improved thermal conduction across interfaces~\cite{gharib2015interfacial, nguyen2021comprehensive}. Moreover, the application of thin metallic or conductive polymer coatings at the interfaces has proven effective in bridging thermal mismatches between different materials~\cite{chen2023recent, zuo2023comparative}.

Additive manufacturing, or 3D printing, complements these nanostructuring and interfacial techniques by enabling the fabrication of composites with complex, tailored thermal properties~\cite{yuan2021additive}. Advanced printing methods, such as coaxial extrusion and the creation of graded materials~\cite{rafiee2021advances,li2020review}, allow for precise control over filler distribution and orientation, crucial for developing effective thermal management solutions like heat exchangers and thermal barriers~\cite{sun2024advanced,favero2021additive}. Additionally, the incorporation of phase-change materials into printed structures expands the potential for high-efficiency thermal energy storage~\cite{freeman2023advanced}, illustrating the versatility of additive manufacturing in the development of next-generation composites tailored for specific thermal management applications~\cite{wang2020fabrication, shemelya2017anisotropy, yuan2021additive}.

\subsection{Computational Advances and Predictive Modeling}
Theoretical and computational methods have made substantial contributions to the understanding and optimization of heat transport in composites. First-principles calculations, MD simulations, numerical methods, and AI-based approaches enable the prediction of thermal properties across micro- to macroscales~\cite{khan2024computational,zha2024polymer}; see Figure~\ref{fig:scale}.
As described in detail in section~\ref{section:modeling}, simulations allow for the investigation of thermal transport mechanisms at the atomic level, providing insight into phonon scattering, interfacial thermal resistance, and the effects of nanostructuring. These simulations help us to understand how molecular interactions and structural configurations influence heat transfer within the composite~\cite{tornabene2021advanced}.
Finite element modeling is employed to simulate heat flow in composites with complex geometries and filler distributions. By modeling the composite structure, it is possible to predict effective thermal conductivity and identify optimal filler arrangements. This approach aids in designing materials with tailored thermal properties for specific applications\cite{tornabene2021advanced}.
Machine learning techniques have been applied to predict thermal properties based on extensive datasets of experimental and simulation results. ML models can identify patterns and correlations that are not readily apparent, assisting in the design of composites with desired thermal properties. These predictive tools accelerate the material design process by reducing the need for exhaustive experimental trials~\cite{mukherjee2024predicting,kielar2024anomalous,luo2023predicting,guo2023fast}.
Advancements in multiscale modeling, combining atomistic simulations with continuum-level approaches, have improved the accuracy of predictions in composite heat transport. This integrated approach allows for the consideration of phenomena occurring at different length scales, from atomic interactions to macroscopic heat flow, providing a comprehensive understanding of thermal behavior in composites~\cite{xu2019modeling}.

\section{Challenges and Future Directions}
\begin{figure}
    \centering
    \includegraphics[width=0.9\linewidth]{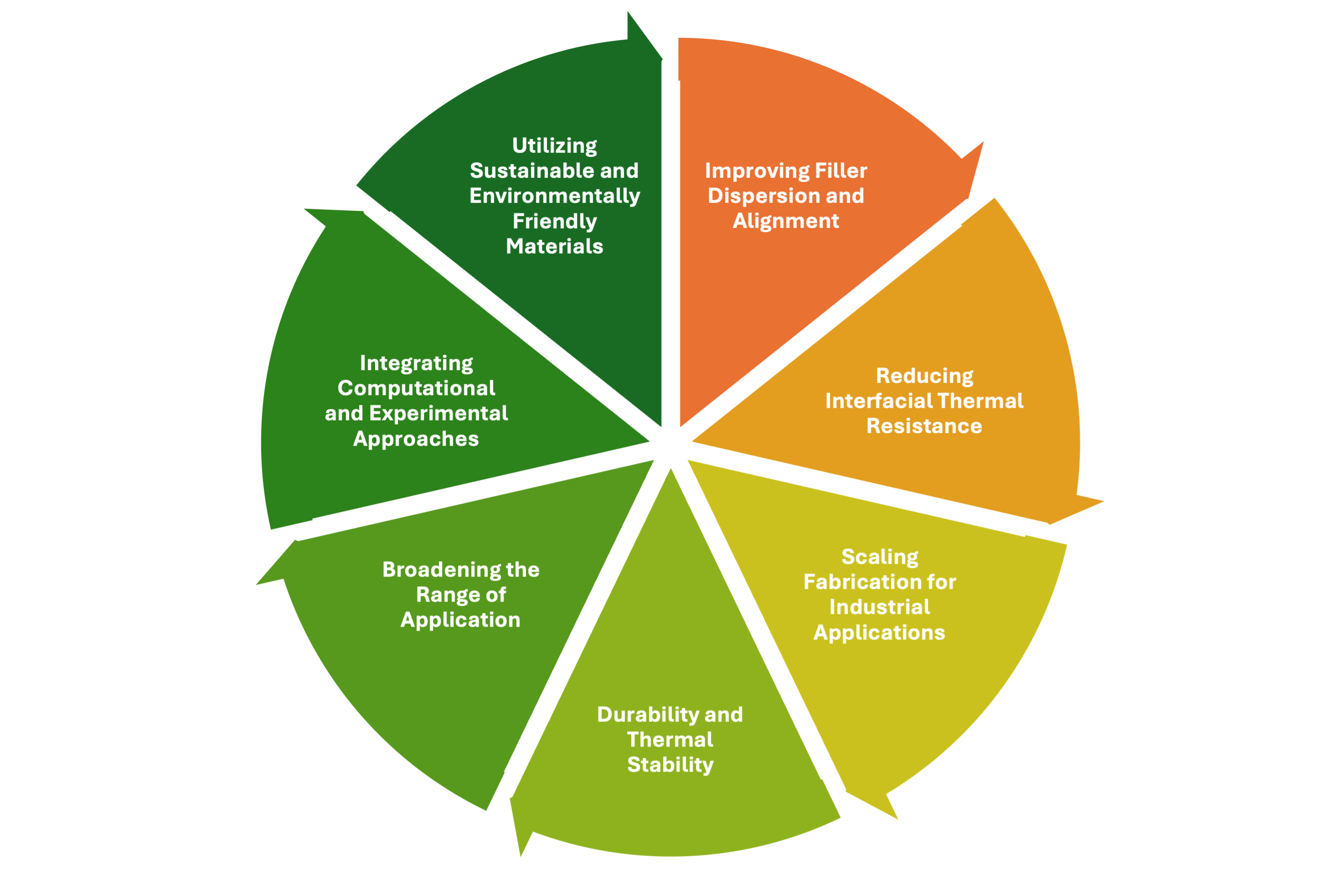}
    \caption{Overview of the primary challenges and future direction in heat transport management of composite materials.}
    \label{fig:challenges}
\end{figure}

Despite substantial advancements in management of heat transport in composite materials, several challenges remain that impede the realization of composites with optimal thermal properties for various applications. Addressing these challenges is essential for the development of next-generation composite materials that meet the complex thermal management demands in electronics, energy, and aerospace industries. This section outlines the primary challenges facing the field, including filler dispersion, interfacial resistance, scalability, durability, and environmental considerations, and suggests future directions that could address these issues and unlock new possibilities in heat transport management, see Figure~\ref{fig:challenges}.

\subsection{Improving Filler Dispersion and Alignment}
Achieving uniform dispersion and precise alignment of high-conductivity fillers such as carbon nanotubes, graphene, and boron nitride nanosheets, within the matrix is a core challenge in composite design. As mentioned before, poor dispersion leads to agglomeration, which disrupts continuous thermal pathways and reduces the overall thermal conductivity of the composite. Agglomerates can also act as stress concentrators, adversely affecting mechanical properties~\cite{xu2021mini}.
Precise alignment of fillers is crucial for maximizing anisotropic heat conduction, which is desirable in applications requiring directed heat flow, such as thermal interface materials and heat spreaders in electronic devices. 
Future research should focus on developing advanced processing techniques to improve filler dispersion and alignment. Techniques such as shear-induced alignment, where shear forces during processing align fillers along the flow direction, offer promising avenues. Field-assisted alignment utilizes external magnetic or electric fields to orient anisotropic fillers like CNTs and graphene sheets. Self-assembly methods leverage the intrinsic interactions between fillers and matrices, allowing fillers to spontaneously organize into ordered structures. Additionally, in-situ growth of fillers within the matrix and functionalization of filler surfaces to improve compatibility with the matrix are strategies that can enhance both dispersion and alignment~\cite{kanu2019self,wang2021ordered,markandan2023fabrication}.
Developing standardized evaluation methods will aid in comparing different composites and processing techniques, facilitating the optimization of materials for enhanced thermal properties.

\subsection{Reducing Interfacial Thermal Resistance}
Interfacial thermal resistance remains a significant barrier to effective heat transport in composites. The mismatch in phonon spectra between fillers and matrix materials leads to phonon scattering at interfaces, limiting heat transfer efficiency across the composite. Addressing ITR requires innovative interfacial engineering strategies~\cite{wei2024innovative,zhong2024recent}.
Computational optimization, employing molecular dynamics simulations and machine learning algorithms, can predict and optimize interfacial properties, guiding the selection of filler-matrix combinations and interface treatments~\cite{wang2024applications,yin2024application}. Developing dynamic and self-healing interfaces, where materials can adapt or repair interfacial bonds under thermal or mechanical stress, presents another promising approach to maintaining high thermal performance over time~\cite{wang2024applications,yu2023machine,alabduljabbar2023modeling}.

\subsection{Scaling Fabrication for Industrial Applications}
Scaling up the production of thermally conductive composites without compromising their tailored properties poses considerable challenges~\cite{mohanty2018composites}. Laboratory-scale techniques often encounter obstacles when adapted for industrial-scale manufacturing due to cost, complexity, and process limitations.
To broaden industrial applicability, efforts should focus on simplifying manufacturing processes and developing cost-effective, scalable techniques. Simplifying manufacturing involves developing straightforward, robust fabrication methods compatible with existing industrial equipment~\cite{astrom2018manufacturing}. Techniques like melt blending and solvent casting can be optimized for better filler integration and uniformity.
Innovations in additive manufacturing, such as fused deposition and direct ink writing, enable the production of complex composite structures with controlled architectures~\cite{astrom2018manufacturing}. Solution processing and roll-to-roll methods offer scalability and efficiency by adapting solution-based processes for continuous production of composite films. Maintaining the intricate composite architectures required for enhanced heat transport during scaling is crucial; process parameters must be carefully controlled to preserve filler dispersion and alignment.
Strengthening collaborations between industry and academia can facilitate technology transfer and address practical manufacturing challenges. Developing standardized protocols and quality control measures will support the reliable production of high-performance composites at scale~\cite{delannoy2022effective}.

\subsection{Durability and Thermal Stability}
Long-term stability and reliability under thermal cycling, mechanical stress, and environmental exposure are crucial for practical applications. Repeated thermal cycling can cause interface degradation, filler detachment, and microstructural changes, diminishing thermal performance and potentially leading to failure~\cite{wang2022thermal,gong2022revealing}.
Future work should concentrate on developing thermally stable materials that retain properties at elevated temperatures~\cite{barra2023comprehensive}. Creating composites with matrices and fillers that maintain integrity under thermal stress is essential. High-temperature polymers and ceramics can enhance thermal stability and extend the operational lifespan of composites.
Incorporating self-healing capabilities into composites can address degradation over time. Self-healing mechanisms, such as microencapsulated healing agents or dynamic covalent bonds, can repair microcracks and interface damage, restoring thermal pathways and mechanical strength. Understanding degradation mechanisms by studying how thermal, mechanical, and environmental factors contribute to composite degradation can guide the design of more resilient materials~\cite{kanu2019self}.
Developing hybrid materials that combine organic and inorganic components can leverage the advantages of both, offering improved thermal and mechanical stability. Protective encapsulation—applying coatings or barriers to protect composites from oxidation, moisture, and chemical exposure—enhances durability in harsh environments. Reliability testing under simulated operational conditions will validate the long-term performance of thermally conductive composites, ensuring they meet the demands of their intended applications~\cite{seno2023multifidelity}.

\subsection{Broadening the Range of Application-Specific Thermal Properties}
Tailoring composites to exhibit specific thermal conductivities is challenging but necessary for diverse applications, ranging from thermal insulation to efficient heat dissipation. Future research directions include developing stimuli-responsive materials that change thermal properties in response to external stimuli such as temperature, light, or electric fields~\cite{shi2018function}. These materials can adapt to varying thermal loads, offering dynamic control over heat flow~\cite{etawy20244d}.
Exploring materials with dynamic thermal properties or switchable thermal conductivities can enable applications like smart textiles or adaptive insulation systems. Expanding material choices by investigating novel fillers and matrices can achieve a wider range of thermal conductivities and functionalities~\cite{du2021wide,zhang2022thermal,babaei2023reversible,zhao2024carbon}.
Designing multi-functional composites that combine thermal management with other properties such as electrical conductivity, mechanical strength, or electromagnetic interference shielding, can meet the demands of advanced applications in electronics, aerospace, and energy systems~\cite{dong2024structure}. Advancements in material science and processing techniques will enable the customization of thermal properties to meet specific application requirements, enhancing the versatility and utility of composite materials.

\subsection{Integrating Computational and Experimental Approaches}
An integrated approach that combines computational predictions with experimental validation accelerates the development of high-performance composites. Improving predictive models involves enhancing the accuracy of computational tools by incorporating realistic material behaviors, including defects, interfacial phenomena, and non-linear responses under different conditions~\cite{mahajan2024optimized,allen2022machine}.
Multiscale modeling, which bridges atomistic simulations with continuum-level analyses, captures phenomena across different length scales, providing a comprehensive understanding of heat transport mechanisms~\cite{fish2021mesoscopic}. High-throughput screening, utilizing machine learning and automation to rapidly evaluate a vast material space, helps identify promising candidates for experimental testing, reducing development time and resource expenditure~\cite{shahzad2024accelerating,athanasiou2024integrating}.
Establishing databases and platforms for sharing experimental and computational data fosters collaboration between researchers, promoting transparency and accelerating progress~\cite{li2023database,huang2022knowledge,lee2022automated}. Validating models experimentally by conducting targeted experiments to verify computational predictions and refining models based on empirical evidence enhances the synergistic approach. This integration enhances understanding, optimizes material design, and leads to more efficient development of composite materials with tailored thermal properties~\cite{le2016discovery,park2024deep,neveu2019gap}.

\subsection{Future Directions in Sustainable and Environmentally Friendly Materials}
Sustainability is becoming increasingly important in material selection and composite design due to growing environmental concerns~\cite{ruggerio2021sustainability}. Future efforts should focus on developing biodegradable and recyclable materials, creating composites using biodegradable polymers and natural fillers like cellulose nanofibers or chitosan to reduce environmental impact~\cite{asyraf2022product,fouad2019design,deng2023sustainable,lunetto2023sustainability}.
Employing green synthesis methods, eco-friendly synthesis routes that minimize hazardous chemicals, energy consumption, and waste generation, contributes to sustainability and reduces the ecological footprint of composite production~\cite{shekar2018green}. Conducting life cycle assessments evaluates the environmental impact of composites throughout their life cycle, from raw material extraction to end-of-life disposal, guiding sustainable design choices~\cite{hermansson2022allocation,bachmann2017environmental}.
Utilizing renewable resources and promoting circular economy principles in composite manufacturing align with environmental goals, ensuring that materials can be reclaimed, recycled, or safely decomposed after use. Ensuring that materials meet environmental regulations and standards facilitates market acceptance and compliance with international guidelines.
Balancing performance with environmental responsibility will be essential for the future of thermally conductive composites~\cite{hu2020nanocellulose,ran2023development,wang2023roadmap,yang2024development}. Integrating sustainability considerations into material design and manufacturing processes will contribute to the development of composites that are not only high-performing but also environmentally friendly.
\section{Conclusion}
Understanding heat transfer in composite materials is essential for optimizing their performance in critical applications across various industries, including aerospace, automotive, electronics, renewable energy, and construction. This comprehensive review has examined the fundamental mechanisms of heat transfer and how they are influenced by the composition and structure of composite materials.

Conductive heat transfer in composites is significantly affected by factors such as reinforcement orientation, geometry, dispersion, interfacial thermal resistance, and the volume fraction of reinforcements. Optimizing these factors enables the design of composites with tailored thermal conductivities, essential for applications requiring efficient heat dissipation or insulation. Convective heat transfer, though less prominent in solid composites, can be enhanced through the incorporation of porosity, internal channels, and engineered surface textures, facilitating improved thermal management in applications like heat exchangers and cooling systems. Radiative heat transfer becomes critical at high temperatures, where surface emissivity and environmental interactions dictate the effectiveness of thermal radiation management.

Modeling and predicting heat transfer in composites utilize a combination of analytical models, numerical methods, atomistic simulations, and artificial intelligence-based models. These approaches provide insights into thermal behavior across different scales, from atomic interactions to macroscopic properties, enabling the optimization of materials for specific thermal performance criteria. Experimental techniques such as steady-state methods, transient techniques, and non-destructive testing are crucial for accurately measuring thermal properties, validating models, and assessing material performance under realistic conditions.

The applications of composite materials in heat transfer are vast and impactful. In the aerospace industry, composites provide thermal protection and structural integrity under extreme conditions. The automotive sector benefits from composites in thermal management systems, enhancing performance and safety. In electronics, composites facilitate efficient heat dissipation, crucial for device reliability and longevity. Renewable energy applications leverage composites for improved efficiency and durability in systems like wind turbines and solar collectors. In construction, composites contribute to energy efficiency and safety through superior insulation and fire-resistant materials.

Recent advances in material design, nanostructuring, interfacial engineering, computational modeling, and the development of emerging materials have significantly enhanced the thermal transport properties of composites. Innovations such as the integration of high-conductivity fillers, precise control over nanostructures, and advanced interfacial treatments have led to composites with enhanced thermal conductivity and tailored properties. Computational tools and machine learning have accelerated material discovery and optimization, enabling the design of composites with specific thermal characteristics.
Despite these advancements, challenges remain. Achieving uniform filler dispersion and alignment, reducing interfacial thermal resistance, scaling fabrication processes for industrial applications, ensuring durability and thermal stability, and addressing environmental sustainability are critical areas that require continued research and development. Future directions include developing advanced processing techniques, innovative interfacial engineering strategies, cost-effective manufacturing methods, and sustainable materials. Integrating computational and experimental approaches will further enhance the understanding and optimization of heat transfer in composites.
The insights provided by this review aim to bridge the gap between fundamental research and practical applications, fostering progress in the design and utilization of composite materials for efficient heat transfer.


\bibliographystyle{unsrt}  
\bibliography{ref}

\end{document}